\documentclass[a4paper,fleqn,usenatbib]{mnras}

\usepackage[T1]{fontenc}
\usepackage{ae,aecompl}
\usepackage{pbox}
\usepackage{color}
\usepackage[utf8]{inputenc}

\usepackage{graphicx}	
\usepackage{amsmath}	
\usepackage{amssymb}	
\usepackage{hyperref}
\usepackage{refcount}
\usepackage{todonotes}
\usepackage [english]{babel}
\usepackage [autostyle, english = american]{csquotes}
\MakeOuterQuote{"}

\def\Mstar{M_{\star}}
\def\MBH{M_{\rm BH}}
\def\Msun{\rm M_{\odot}}
\def\log10{\rm log_{10}}

\title[Black hole monster galaxies in cluster environments]{Galaxies with monstrous black holes in galaxy cluster environments }

\author[L. van Son et al.]{Lieke A. C. van Son,$^{1}$\thanks{E-mail: vson@strw.leidenuniv.nl}
Christopher Barber,$^{1}$
Yannick M. Bah\'{e},$^{1}$
\newauthor
Joop Schaye,$^{1}$
David J. Barnes,$^{4,5}$
Robert A. Crain,$^{2}$
\newauthor 
Scott T. Kay,$^{5}$
Tom Theuns,$^{3}$
Claudio Dalla Vecchia$^{6,7}$ 
\\ 
$^{1}$Leiden University, Niels Bohrweg 2, 2333 CA Leiden, The Netherlands\\
$^{2}$Astrophysics Research Institute, Liverpool John Moores University, 146 Brownlow Hill, Liverpool L3 5RF, UK \\
$^{3}$Institute for Computational Cosmology, Department of Physics, University of Durham, South Road, Durham DH1 3LE, UK \\
$^{4}$ Department of Physics, Kavli Institute for Astrophysics and Space Research, Massachusetts Institute of Technology, \\ Cambridge, MA 02139, USA \\
$^{5}$Jodrell Bank Centre for Astrophysics, School of Physics and Astronomy, The University of Manchester, Manchester M13 9PL, UK\\
$^{6}$Instituto de Astrof\'{i}sica de Canarias, C/V\'{i}a L\'{a}ctea s/n, 38205 La Laguna, Tenerife, Spain \\
$^{7}$Departamento de Astrof\'{i}sica, Universidad de La Laguna, Av. del Astrof\'{i}sico Franciso S\'{a}nchez s/n, 38206 La Laguna, Tenerife, Spain
}

\date{Accepted XXX. Received YYY; in original form ZZZ}

\pubyear{2017}

\begin{document}
\label{firstpage}
\pagerange{\pageref{firstpage}--\pageref{lastpage}}
\maketitle

\defcitealias{Barber2016}{B16}

\begin{abstract}
Massive early-type galaxies follow a tight relation between the mass of their central supermassive black hole ($\MBH$) and their stellar mass ($\Mstar$). The origin of observed positive outliers from this relation with extremely high $\MBH$ ($> 10^{9} \Msun$) remains unclear. We present a study of such outliers in the Hydrangea/C-EAGLE cosmological hydrodynamical simulations, designed to enable the study of high-mass galaxy formation and evolution in cluster environments. We find 69 $M_{\rm BH}(M_{\star})$ outliers at $z=0$, defined as those with $\MBH >10^{7} \Msun$ and $\MBH/\Mstar > 0.01$. This paper focusses on a sample of 5 extreme outliers, that have been selected based on their $\MBH$ and $\Mstar$ values, which are comparable to the most recent estimates of observed positive outliers. This sample of 5 outliers, classified as ``Black hole monster galaxies'' (BMGs), was traced back in time to study their origin and evolution. In agreement with the results of previous simulations for lower-mass $\MBH(\Mstar)$ outliers, we find that these galaxies became outliers due to a combination of their early formation times and tidal stripping. For BMGs with $\MBH > 10^9 \Msun$, major mergers (with a stellar mass ratio of $\mu > 0.25$) at early times ($z>2$) precede the rapid growth of their supermassive BHs. Furthermore, the scatter in the relation between $\MBH$ and stellar velocity dispersion, $\sigma$, correlates positively with the scatter in [Mg/Fe]($\sigma$). This indicates that the alpha enhancement of these galaxies, which is closely related to their star formation history, is related to the growth of their central BHs. 
\end{abstract}


\section{Introduction}

The theory of galaxy formation is one of the most active topics of modern-day cosmology. To aid our understanding of the physical processes involved, many empirical scaling relations between galaxy properties have been deduced. One of the strongest of such relations is that between the mass of a supermassive black hole (BH) and the stellar mass of its host galaxy, $M_{\star}$ \citep[for a review on supermassive BH scaling relations, see][]{KormendyHo2013}. This strong correlation hints at a co-evolution between the central BH and the galaxy as a whole. 

Recently, a handful of outliers to this relation have been observed, the most interesting being those with extremely high black hole mass ($M_{\rm BH}$) for their stellar mass. Well-studied examples are NGC 1271 \citep{ExplainNGC1271,ObsNGC1271}, NGC 1277 \citep{NGC1277.Bosch2012, NGC1277Clust2014, ObsNGC1277}, and Mrk 1216 \citep{Mrk1216andPGC032873from2017}. The origin (and even the existence) of such over-massive BHs is highly debated. For example, \cite{ExplainReportedlyOvermassive} and \cite{ExplainNGC1271} both argue that when spheroid/disc decompositions are modelled under different assumptions, the $M_{\rm BH}$ values of e.g. NGC 1277 and NGC 1271 can be brought to within 2-$\sigma$ of the observed $M_{\rm BH}(M_{\star})$ relation, and are no longer anomalous. It is therefore interesting to test whether such $M_{\rm BH}(M_{\star})$ outliers are expected to exist in current galaxy formation theories, which can most accurately be tested via cosmological simulations. 

\citet[hereafter B16]{Barber2016} studied positive outliers of the $M_{\rm BH}(M_{\star})$ relation in the (100 Mpc)$^3$-volume EAGLE simulation \citep{TheEaglePaper,Crain2015}. They confirmed that such outliers do indeed exist, and arise due to their early formation time and/or the tidal stripping of stars. The importance of tidal stripping was also pointed out by \cite{Volonteri2008} and \cite{Volonteri2016}. However, \cite{Barber2016} did not find outliers with $M_{\rm BH} > 10^{8.5} \rm M_{\sun}$ although, as mentioned above, such BH outliers have been observed. 
Plausibly, the lack of high-mass outliers in \citetalias{Barber2016}, is due to the limited volume of the EAGLE simulation, as \cite{Saulder2015} found that the frequency of massive compact high stellar velocity dispersion galaxies (such as NGC1271 and NGC1271) is $\sim 10^{-7}$ galaxies per $\rm Mpc^{-3}$, implying that $<1$ of such galaxies should be found in the (100 Mpc)$^3$ volume studied by \citetalias{Barber2016}. A larger simulation volume would thus be required to find $M_{\rm BH}(M_{\star})$ outliers with such high $M_{\rm BH}$. In addition, simulations of representative volumes, such as EAGLE, are dominated by field galaxies, while we expect to find more high-mass BHs (and thus also $\MBH(\Mstar)$ outliers with $\MBH$ of > $10^{8.5} \rm M_{\sun}$) in galaxy cluster environments due to the higher densities and galaxy masses. Moreover, tidal stripping is expected to be enhanced in such galaxy cluster environments due to the stronger tidal forces.

The Cluster-EAGLE (C-EAGLE) project \citep{Hydrangea,C-EAGLE}, a suite of hydrodynamical, cosmological zoom-in simulations of galaxy clusters and their surrounding environments, is ideal for the purpose of this research. 
This work aims to study the origin and properties of the $M_{\rm BH}(M_{\star})$ outliers in the cluster environments of the Hydrangea simulations \citep[a subset of the C-EAGLE simulations, for which the high-resolution zoom-in region extends to 10 times the virial radius]{Hydrangea} in a stellar mass regime closer to that of observed galaxies with overly-massive BHs. We compare our results to those from \citetalias{Barber2016} for the (100 Mpc)$^3$ EAGLE volume and investigate the physical origin of galaxies with extreme BH masses. 
When outliers with $\MBH > 10^{8.5} \Msun$ are found in Hydrangea, it is interesting to see whether the origin of their anomalous $\MBH(\Mstar)$ could be explained by the same arguments as \citetalias{Barber2016}, or whether additional reasoning is required. Recently, \cite{McAlpine2018} and \cite{Steinborn2018} investigated the effect of major mergers on triggering the rapid growth phase of central black holes. Major mergers are defined as $M_{\star,1}/M_{\star,2} = \mu > 0.25$, while intermediate mergers are $ 0.1 < \mu < 0.25$, where $M_{\star,1}$ and $M_{\star,2}$ are the the stellar masses of the less and more massive progenitor of the galaxy in question. They both argue that major galaxy-galaxy interactions are not statistically dominant mechanisms for fuelling nuclear activity or triggering the rapid growth phase at $z > 2$. We therefore investigate whether our most extremely positive outliers (with $\MBH > 10^{9} \Msun$) have experienced any major mergers.

This paper is organized as follows: we discuss the Hydrangea simulations and our galaxy tracing methods in Section \ref{sec: methods}. The simulated $\MBH(\Mstar)$ distribution at $z=0$ is evaluated in Section \ref{sec: outliers at z =0}.  With the purpose of comparing the simulated BMGs to recently observed outliers, we continue with a more detailed evaluation of the 5 most interesting BMGs in Section \ref{sec: UMBH evolution}, where we also consider the effects of tidal stripping and early formation time on $\MBH(\Mstar)$ outliers in Section \ref{ss: tidal stripping formation time}. We conclude with a short discussion on the [Mg/Fe] ratio of galaxies in Hydrangea (subsection \ref{ss: Mg/Fe}). Our main findings are summarized in Section \ref{sec: conclusion}.

Throughout this paper, the following cosmological parameters, consistent with \cite{TheEaglePaper} and \cite{2014planck} are assumed: $\Omega_{\rm m}$ = 0.307, $\Omega_{\rm b}$ = 0.04825, $\Omega_{\Lambda}$ = 0.693, $\sigma_{8}$ = 0.8344, and  $ h = \rm H_0$ /(100  km s$^{-1}$  Mpc$^{-1}$) $= 0.6777$. 

\section{Simulations and Analysis}
\label{sec: methods}

\subsection{The Hydrangea simulations}
\label{ss: data}
The EAGLE project is a suite of hydrodynamical simulations aimed at understanding how galaxies form and evolve from shortly after the Big Bang to the present day \citep{TheEaglePaper,Crain2015}. The public release is described in \citet{MergerTreesEAGLE}.  
The C-EAGLE (short for Cluster-EAGLE) project \citep{C-EAGLE,Hydrangea} is a suite of cosmological hydrodynamical zoom-in simulations that adopts the EAGLE galaxy formation model to study the formation and evolution of 30 galaxy clusters with masses ranging from\footnote{$M_{200}$ is defined as the mass within $r_{200}$, which is the the radius at which the mean enclosed mass density $\rho$ is equal to 200 times the critical density ($\rho_{\rm crit} = 3H^{2}/8\pi G$) of the Universe.} $M_{200} = 10^{14}$ to $\approx 2 \times 10^{15} \Msun$. 
The Hydrangea project \citep{Hydrangea} is a core part of the C-EAGLE project, which consists of 24 of the 30 total galaxy clusters, simulated with high-resolution zoom-in regions extending out to 10$r_{200}$ from each cluster's centre. 
Particles in the re-simulated higher-resolution regions initially have masses identical to those in the intermediate-resolution EAGLE simulations: $1.81 \times 10^{6} \Msun$ for gas particles and  $9.7 \times 10^{6} \Msun$ for dark matter particles.

Particle information was saved in regular time step intervals of 500 Myr in 28 snapshots, running from $z = 14.0$ to $z = 0$. Additionally, snapshots 19 ($z = 0.37$) and 26 ($z = 0.1$) were created to enable comparison to the EAGLE simulation.  
Furthermore, between each of the 28 main snapshots, three ‘snipshots’ were stored, reducing the time resolution to $t = 125 \rm Myr$. These snipshots contain only the most important, and most rapidly time-varying, quantities, such as particle positions and velocities \citep[similar to EAGLE; see][]{TheEaglePaper}. For the 1 Gyr intervals at lookback times of 0–1, 5–6, and 8–9 Gyr, 19 additional snipshots were stored, boosting the time resolution to $t = 25 \rm Myr$ for those epochs.

A full description of the model is given by \cite{TheEaglePaper}, while a complete overview of the C-EAGLE project and Hydrangea simulations is given by \cite{C-EAGLE} and \cite{Hydrangea}, respectively.  Here we will summarize the aspects that are most relevant for our study.

Star particles are formed stochastically from gas particles once their density exceeds the metallicity-dependent threshold of \cite{Schaye2004StarForm}. The star formation rate is obtained by converting the Kennicutt-Schmidt law into a pressure law \citep{VecchiaSchaye2008}, and stellar mass loss from massive stars, AGB stars and type Ia and type II supernovae is simulated by the redistribution of stellar mass and metals among neighbouring gas particles \citep{Wiersma2009b}. The mass of the individual elements lost through stellar feedback processes (i.e. winds from AGB stars,  massive  stars,  and  core  collapse  supernovae) is computed using the nucleosynthetic yields and metallicity-dependent lifetimes of \cite{Marigo2001} and \cite{Portinari1998}. Stellar energetic feedback is implemented by injecting thermal energy into the neighbouring gas particles, following \cite{VecchiaSchaye2012}. During each stochastic feedback event, the temperature increase of the receiving gas is kept fixed, while the number of influenced gas particles depends on the local gas density and metallicity, and was calibrated to reproduce the observed $z=0.1$ galaxy stellar mass function and galaxy sizes \citep{Crain2015}. 
The elements most important for radiative cooling \citep[H, He, C, N, O, Ne,  Mg,  Si,  and  Fe;][]{Wiersma2009a}  are  traced  individually. Metallicities mentioned in this work refer to stellar metallicity. For SN Type Ia, yields are taken from the W7 model of \cite{Thielemann2003}.

Following \cite{SpringelBHs2005}, a BH is seeded in a Friends of Friends (FOF) halo that does not already contain a BH, once its total mass exceeds $10^{10} \Msun$ $h^{-1}$. The gas particle with the highest density is converted into a BH particle with an initial sub-grid seed mass of $10^{5} \Msun$ $h^{-1}$. $M_{\rm BH}$ can subsequently grow through gas accretion and BH-BH merging \citep{TheEaglePaper,BHgrowthYetli2016}. The accretion model is based on a sub-grid accretion disc with an adjustable viscosity parameter, $C_{\rm visc}$, that influences the $\MBH(\Mstar)$ relation \citep{Crain2015}, and was calibrated to reproduce the galaxy stellar mass function (GSMF). To model (unresolved) dynamical friction effects, BHs with $M_{\rm BH}$ < 100 $m_{\rm gas}$ are forced to migrate towards the minimum of the gravitational potential of the halo.

BHs merge when they are spatially separated by less than the BH smoothing kernel and three times the gravitational softening length, and their relative velocity is smaller than the circular velocity at the smoothing length of the most massive BH of the pair. The latter condition prevents BHs from merging during flybys. 

C-EAGLE uses EAGLE's `AGNdT9' model to simulate AGN feedback, which differs from the reference model used in EAGLE only by a different value of  $\Delta T_{\rm AGN}$ (= $10^9$ K instead of $10^{8.5}$ in EAGLE) and $C_{\rm visc}$(= $2\pi \times 10^{2}$), since this produces more realistic hot gas haloes of galaxy groups. This reduces numerical cooling losses but still results in somewhat too large hot gas mass fractions in clusters \citep{C-EAGLE}. 
AGN-feedback is implemented by stochastically heating the surrounding gas particles, where the stochastic probability is chosen to ensure that on average the energy injection rate is $\epsilon_{r} \epsilon_{f} \dot{m}_{\rm acc,BH} c^2$, where $\epsilon_{r} = 0.1$ is the radiative efficiency of the accretion disc and $\epsilon_{f} = 0.15$ is the fraction of the radiated energy that is coupled to the interstellar medium. The efficiency of BH feedback was calibrated to yield a match to the observed $M_{\rm BH}(M_{\star})$ relation at $z \approx 0$.

\subsection{Galaxy identification and corrections}
\label{ss: groups and subgroups}
Dark matter haloes are identified using the FOF algorithm on dark matter particles with linking length $0.2$ times the mean inter-particle spacing \citep{FOF}. 
Baryon particles are then attached to the halo of the nearest DM particle (if this DM particle belongs to any halo). The SUBFIND algorithm \citep{Springel2001, Dolag2009} is used subsequently to identify gravitationally self-bound substructures within these FOF haloes, i.e. objects that can be identified as galaxies. Subhaloes of fewer than 20 particles are not catalogued as they are poorly sampled. 
The subhalo in the FOF-group that contains the particle with the deepest gravitational potential is defined as the central galaxy of that group, the other subhaloes are labelled as satellites. We define a subhalo's stellar mass, $\Mstar$, as the sum over all stellar particles bound to the subhalo. \footnote{When using $M_{\star, 30}$, defined as the sum over all stellar particles bound to the subhalo which are within a 30 proper kpc aperture centred on the potential minimum, the simulations are consistent with the observational definition of stellar mass based on Petrosian radii \citep{TheEaglePaper,Furlong2015}. For the majority of the extreme $M_{\rm BH}$ outliers studied in this work, $M_{\star, 30}$  and $M_{\star}$ are equal at $z= 0$. The maximum difference encountered was $\rm log_{10} \Mstar/\Msun$ - $\rm log_{10} M_{\star, 30}/\Msun$ $\lesssim 10^{-2}$.} To avoid contamination from low resolution regions of the simulation, we only include galaxies which have always been > 2 cMpc away from a low resolution boundary particle at $z=0$.

\subsubsection{Subhalo corrections}
\label{sss: corrections}
Since SUBFIND identifies galaxies as bound structures in the simulation, it can occur that the central BH in a subhalo temporarily becomes its own bound structure with a small fraction of the stars in the central region. This results in subhaloes seeming to spuriously pop in and out of existence over time \citepalias{Barber2016}. While such structures may indeed be self-bound, they would not be identified as individual galaxies in observations, so it is crucial to this work to remove such structures from the data by merging them back into the surrounding galaxy. As in \cite{TheEaglePaper}, subhaloes separated by a distance less than 3 proper kpc and smaller than the half-mass radius of the star particles were merged in post-processing. 

The SUBFIND algorithm is known to occasionally assign BHs to the host of a satellite galaxy instead of to the galaxy in whose centre it resides \citepalias{Barber2016}. In this case, the satellite is typically left without any black hole particles, causing its $M_{\rm BH}$ to temporarily drop to zero, typically during pericentric passages while orbiting its central. We correct for this BH misassignment following \citetalias{Barber2016}. We select candidate subhaloes for each BH particle at each snapshot, at a distance $d$ smaller than the half-mass radius of the relevant subhalo and $ d < 3$ kpc (in proper distance). If the BH was not part of the sub-group of one of these candidates, we selected the most massive candidate as the correct subhalo for the BH and reassigned the particle to that subhalo. This reassignment affected $\approx 0.5 -1 \%$ of the BH particles per snapshot, but had no effect on the final sample of BMG-galaxies.

\subsubsection{BMG selection}
\label{sss: BMG selection}
In order to define (positive) $M_{\rm BH}(M_{\star})$ outliers as galaxies with a central SMBH that is much more massive (or equivalently have a stellar mass that is much lower) than expected relative to the median $M_{\rm BH} - M_{\star}$ relation at $z = 0$, we select galaxies with $M_{\rm BH}/M_{\star} > 0.01$ and $M_{\rm BH} > 10^{7} \Msun$. The former criterion was chosen to select only the strongest outliers in the simulation, while the latter avoids resolution effects caused by seed mass ($10^{5} h^{-1} M_{\sun}$) BHs. This definition differs from that of \citetalias{Barber2016}, who defined outliers as having $\MBH$ more than 1.5 dex above the median $\MBH(\Mstar)$ relation in the simulation. 
When using the definition from \citetalias{Barber2016}, the fraction of all galaxies with $\MBH > 10^{7} \Msun$ at $z=0$ that would be classified at outliers in EAGLE and Hydrangea respectively amounts to $0.075\%$ and $10\%$. Our new definition of being an outlier corresponds to $0.035\%$ and $1.8\%$ of the galaxies with $\MBH > 10^{7} \Msun$ at $z=0$ for EAGLE and Hydrangea respectively. This new definition thus prevents a large fraction of galaxies from being classified as outliers, while still including the galaxies with high $\MBH$ (> $10^{9} \Msun$). For comparison, the value of the median $M_{\rm BH}/M_{\star} \approx$  0.001 at $z= 0$ for $M_{\star} \approx 10^{11} \Msun$.

Since our new definition still results in quite a large set (69) of $M_{\rm BH}(M_{\star})$ outliers, we manually select the most extreme positive outliers of interest for further analysis, i.e. those with the highest ratio of $\MBH/\Mstar$. 
We selected 16 of the most extreme outliers of the $M_{\rm BH}(M_{\star})$ relation, based on a visual inspection of their locations in the $M_{\rm BH}-M_\star$ plane (see Figure \ref{fig: dist}), which will hereafter be referred to as ``Black hole monster galaxies'' (BMGs) \citep[using the term first coined by][]{KormendyHo2013}. 

The stellar mass surface density images of the main progenitors of each of these BMGs were carefully inspected at each snapshot by means of the merger tree algorithm as described below in Section \ref{sss: tracing methods}. Additionally, the stellar surface mass density images of the region around each main progenitor were studied at each output, exploiting the high time resolution of the combined snap- and snipshots.

\subsubsection{Tracing the main progenitors}
\label{sss: tracing methods}
The main progenitors of these ``BMGs'' were primarily traced using the ``merger tree'' algorithm from  Bah\'{e} et al. (in prep.), which is similar to that used for EAGLE \citep{MergerTreesEAGLE,MergerTrees.Qu2017}. In essence, this algorithm identifies candidate descendants of a subhalo in the subsequent snapshot by following its five per cent most bound collisionless particles (i.e. DM, stars, and BHs) as tracers. The actual descendant subhalo is then chosen according to the fraction of tracer particles it contains. In the case of multiple candidate progenitors for the same descendant, the descendant subhalo is chosen according to the total mass contributed by each of them.

The BMGs typically experience complex interactions with their environment, making it difficult to trace their progenitors between snapshots by means of a single algorithm. Hence we confirmed the main progenitors based on an additional visual analysis of the stellar density profiles. 
The $\Mstar$, $\MBH$ and $M_{\rm DM}$ surface density profiles of the BMGs and their environment were visualized at each output (including snipshots), so that they could be subjected to careful inspection by eye (see Figure \ref{fig: Outlier E} for a smoothed example of such an image). 
For some complicated events, the tracing algorithm did not yield the correct progenitor masses. For example, this can occur if the BH particle becomes unbound between snapshots, or when a large fraction of the stellar or DM particles was temporarily (for one or two snapshots) bound to a more massive neighbouring galaxy.
A manual inspection of the stellar density, gas density and the BH particles of the region (of up to $\sim$150 pkpc) surrounding the subhalo in question was therefore decisive in determining the correct progenitor mass in such cases. 

It was found that 9 of the 16 BMG-subhaloes lie very close (at a distance $d < 15$ pkpc) to a more massive neighbour at $z = 0$. These 9 subhaloes fell just outside of our merger and BH reassignment requirements of $d < 3$ pkpc, but were clearly an artefact of the SUBFIND algorithm or in the final stages of a galaxy-galaxy merger as described above. One additional subhalo was classified as being in the final stages of a merging event between two massive ($\Mstar \sim 10^{11}$ M$_{\sun}$) galaxies. These high masses have a larger gravitational influence radius and therefore reside at a larger distance (>50 kpc) from its merging companion.
Lastly, one subhalo was found to consist of only 23 baryonic particles at $z = 0$, where there was no significant stellar over-density at the location of this subhalo. It was therefore classified as a naked/ejected BH without a real galaxy hosting it.

We concluded that these 11 subhaloes would not be considered an BMG in the observational sense and should therefore not be considered as individual galaxies. We thus excluded them from further analysis, and focus on the remaining 5 BMGs. \footnote{Videos of the stellar density profiles, including an extended description of each of the initial sample of 16 subhaloes can be found at \url{https://sites.google.com/view/vson-umbh-galaxies/homepage} }

\section{Resulting $\MBH(\Mstar)$ Outliers at $z=0$}
\label{sec: outliers at z =0} 
\begin{figure*}
    \centering
	\includegraphics[trim=3.0cm 1cm 10.0cm 5.5cm, clip=true, width= 0.9\textwidth]{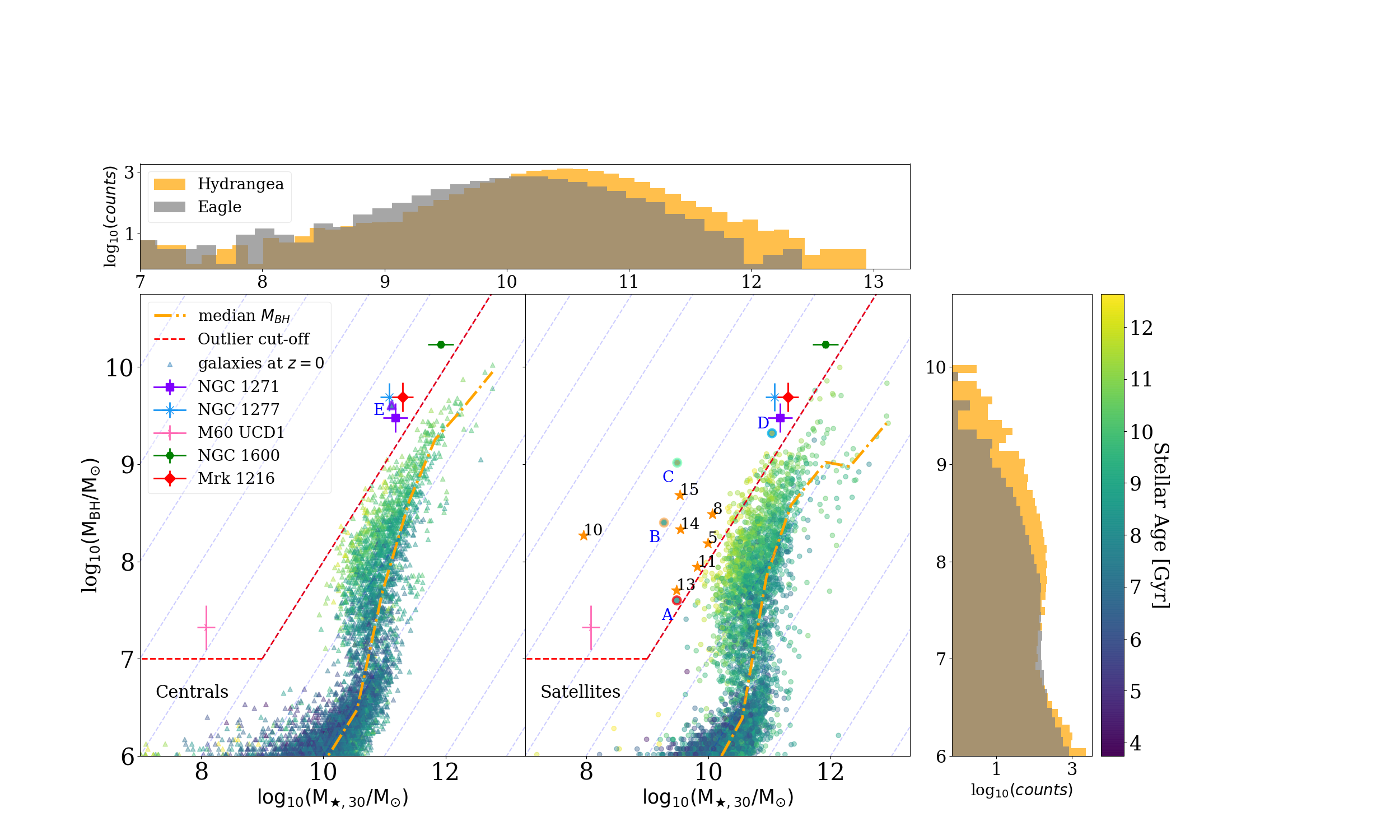}
\caption{Galaxy black hole mass as a function of stellar mass for Hydrangea galaxies at $z = 0$. Circular markers denote satellite galaxies (lower middle panel), whereas triangles correspond to central galaxies (lower left panel). Colours (ranging from blue to yellow) indicate the mean mass-weighted age of the stars in Gyr. The limit above which galaxies are defined as $\MBH(\Mstar)$ outliers ($M_{\rm BH}/M_{\star}$ > 0.01 and $M_{\rm BH}$ > $10^{7} \Msun$) is shown as a dashed red line. The median $M_{\rm BH}$ in bins of $M_{\star}$ is shown as a dash-dotted orange line. We also indicate the 5 'Black hole Monster Galaxies' (BMGs) labelled with their BMG-ID (see Table \ref{tab: extreme outliers}). Galaxies from the (100 Mpc)$^3$ EAGLE simulation that match our definition of $\MBH(\Mstar)$-outlier are shown by star-shaped markers and labelled with numbers matching the outlier IDs from \protect\citetalias{Barber2016}. Values for several observed $M_{\rm BH}(M_{\star})$ outliers are shown as points with error bars, their values are taken from \protect\cite{ObsNGC1271}, \protect\cite{ObsNGC1277}, \protect\cite{ObsDwarfGal2014}, \protect\cite{FerreMateu2015} and \protect\cite{Mrk1216andPGC032873from2017}. Histograms of the $M_{\star}$ and $M_{\rm BH}$ distributions for Hydrangea (orange) and the (100 Mpc)$^3$ EAGLE simulation (gray) are shown along the axes; here bin-sizes of 0.15 dex in $M_{\star}$ and 0.2 dex in $M_{\rm BH}$ are used. We predict the existence of galaxies in cluster environments that are strong $M_{\rm BH}(M_{\star})$ outliers with $M_{\rm BH}$ up to $4 \times 10^{9} \Msun$, comparable to the most massive BH outliers observed. \label{fig: dist}}
\end{figure*}

Figure \ref{fig: dist} shows the $M_{\rm BH}-M_\star$ plane for Hydrangea central and satellite galaxies.  A red dashed line marks our threshold for selecting    $M_{\rm BH}(M_\star)$ outliers as described in Section \ref{sss: BMG selection}. At $z = 0$, 77$\%$ of the $M_{\rm BH}(M_{\star})$ outliers are satellites and typically have $M_{\star} = [10^{9} - 10^{11}] \Msun$. The median relation, shown as a solid red line, is lower for satellite galaxies than the median relation for centrals at $\Mstar \approx 10^{12} \Msun$. This is caused by a small subset ($\sim 20$) of `undermassive' BH galaxies (with $\MBH \approx 10^{8.5} \Msun$) that are most likely caused by the ejection of black holes through tidal encounters of similar-mass black holes. Throughout this paper, we compare outliers to the total median $\MBH(\Mstar)$ for both centrals and satellites.

The 5 manually selected BMGs are labelled in Figure \ref{fig: dist} with IDs corresponding to those in table \ref{tab: extreme outliers}. They were selected with a preference for a high $M_{\rm BH}$ because of the similarity to the extremely high $\MBH$ of several observed $\MBH(\Mstar)$ outliers. A bias towards higher $M_{\rm BH}$ at fixed $\Mstar$ also leads to a bias towards older galaxies; as the gradient in stellar age in Figure \ref{fig: dist} shows, at fixed $M_\star$, galaxies with higher $M_{\rm BH}$ tend to be older. There is a larger variation in $\Mstar$ at fixed $\MBH$ for satellite galaxies, with respect to that for centrals at $\Mstar \sim 10^{10} \Msun$. This could be an indicator of tidal stripping in such cluster environments. However, it could also plausibly be caused by the quenching of star formation in satellites as described by e.g. \cite{Hydrangea}. 

In Figure \ref{fig: dist}, we also show histograms of the $M_{\rm BH}$ and $M_{\star}$ distributions of both the (100 Mpc)$^3$ EAGLE volume and the Hydrangea simulations. Unsurprisingly, the Hydrangea simulations contain more galaxies than EAGLE at high $M_{\star}$, and therefore also of supermassive BHs. Because of this larger number of massive galaxies, they are also more likely to contain $\MBH(\Mstar)$ outliers at this high-mass end. 
Outliers with the highest $M_{\rm BH}$ (outliers D and E) have masses comparable to the most recent estimates for NGC 1277 and NGC 1271 \citep{ObsNGC1277, ObsNGC1271}. For comparison, the outliers identified by \citetalias{Barber2016} in the (100 Mpc)$^3$ EAGLE simulations that satisfy our requirements of being a $M_{\rm BH}(M_{\star})$ outlier are shown in Figure \ref{fig: dist} by star-shaped symbols and are labelled with the outlier IDs from \citetalias{Barber2016}. We remind the reader that \citetalias{Barber2016} identified outliers as galaxies that lie 1.5 dex above the median $M_{\rm BH}(M_{\star})$ relation in the simulation at $z=0$, which differs from our threshold (see Section \ref{sss: BMG selection}).

The distribution of galaxies in the $M_{\rm BH} - M_{\star}$ plane can be divided into two general regions. As described by \cite{RapidGrowthPhase}, above a halo mass of $10^{12} \Msun$, stellar feedback is no longer sufficient to prevent gas infall towards the central BH once $M_{\star}$ reaches $\sim$ $10^{10.5} \Msun$. The BH subsequently begins a phase of runaway growth during which it accretes gas until it is massive enough to regulate its own growth \citep[this process is described in more detail in ][and in Section \ref{ss: tidal stripping formation time}]{2017Weinberger,RapidGrowthPhase}. The rapid growth phase causes a nearly vertical trend at $M_{\star}$ $\approx$ $10^{10.5} \Msun$. Once the BH is massive enough to regulate its own growth via AGN feedback, the galaxy follows the shallower high-mass 1:1 relation (i.e. parallel to the gray dashed lines), where growth in $\Mstar$ and $\MBH$ is increasingly due to mergers.

As an extension to the work done by \citetalias{Barber2016}, we find from the $M_{\rm BH} - M_{\star}$ distribution in Figure \ref{fig: dist} that strong $M_{\rm BH}(M_{\star})$ outliers with $M_{\rm BH}$ greater than $10^{9} \Msun$ are expected to exist in cluster environments. These high values are comparable to the most massive observed $\MBH(\Mstar)$ outliers. In the next Section we will study the origin of these BMGs, and compare this to results from \citetalias{Barber2016}.

\section{Evolution and origin of black hole monster galaxies}
\label{sec: UMBH evolution}
As described in Section \ref{sss: BMG selection}, we manually select 5 outliers of interest based on a visual inspection of their locations in the $M_{\rm BH}-M_\star$ plane (see Figure \ref{fig: dist}). These 5 outliers, or BMGs, were selected based on their extreme BH masses and similarity to observed $\MBH$ outliers. We have traced their main progenitors back in time to study their origin and evolution. 

\begin{figure}
    \centering
    \includegraphics[width= 0.45\textwidth]{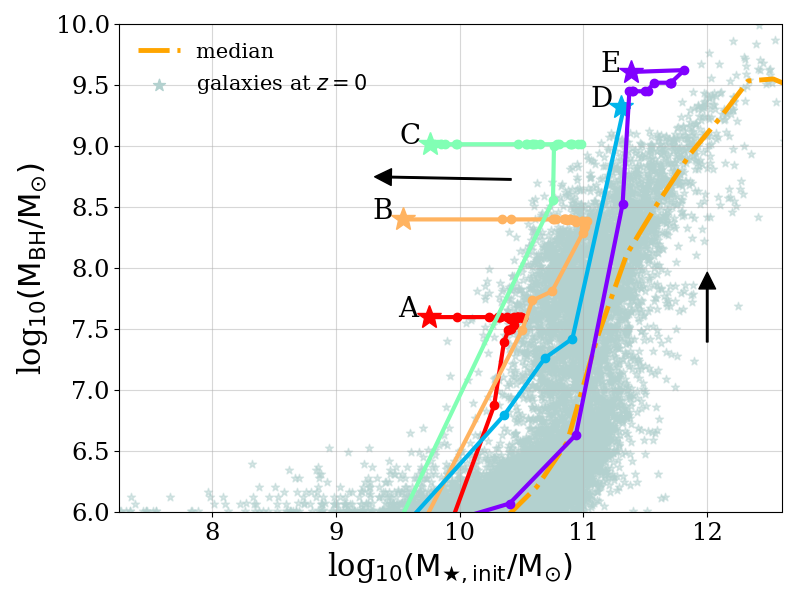} 
\caption{Evolutionary tracks of the main progenitors of the BMGs in the $\MBH$-$\rm M_{\star,\rm init}$ plane, where $ M_{\star,\rm init}$ is defined as the the stellar mass using the mass of each star particle at birth. Individual snapshots are indicated with circles, and the final snapshot (corresponding to $z = 0$) is marked with a star. The distribution for all galaxies at $z = 0$ is shown in gray-blue. The median of the Hydrangea sample is plotted as a dash-dotted orange line. The typical movement of BMGs in the $\MBH-\Mstar$ plane over time is indicated by black arrows. \label{fig: evolution}}
\end{figure}

In Figure \ref{fig: evolution} we show the evolution of the main progenitors for the five BMGs in the $\MBH-\Mstar$$_{\rm init}$ plane, where 
$M_{\star,\rm init}$, the initial stellar mass, is the the total stellar mass using the mass of each star particle at birth, thereby excluding mass loss due to stellar evolution. Individual snapshots are indicated with circles. The typical movement of BMGs in the diagram over time is indicated by black arrows and the last snapshot ($z=0$) is indicated with a star. These tracks were constructed by examining the merger trees and the stellar mass density profiles of all BMGs, as described in Section \ref{sss: tracing methods}. 

\subsection{Tidal stripping and early formation}
\label{ss: tidal stripping formation time}
We would like to know whether the physical mechanisms by which these galaxies become such extreme outliers of the $M_{\rm BH}(M_{\star})$ relation are consistent with those found by \citetalias{Barber2016}. We therefore investigate the relative contributions of tidal stripping and early formation time to their $\MBH/\Mstar$ ratio.

The outliers are in general relatively compact with respect to the complete sample of galaxies at $z=0$. That is, $78\%$ of the outliers have a stellar half-mass radius below the median at the same $\Mstar$. Furthermore, we compared the distances of satellite outliers to their nearest more massive neighbour. The outliers lie significantly closer to their more massive neighbours than other satellites with $\Mstar > 10^{9} \Msun$, as shown by a KS-test with $p \ll 0.01$. These results point towards tidal stripping as an important contributor in the process of galaxies becoming outliers. 

To investigate the extent to which stripping of stellar material due to tidal forces has influenced our BMGs, we compute the maximum initial stellar mass ($M_{\star,\rm init, max}$) that each BMG reached during its evolution. The unbinding of star particles is the only effect that causes galaxies to lose initial stellar mass, making the difference between $M_{\rm \star,\rm init,max}$ and the $M_{\star, \rm init}$ (at $z=0$) a good measure of stellar stripping. Following \citetalias{Barber2016}, we quantify the impact of stripping by the logarithmic ratio of $M_{\star,\rm init,max}$ and $M_{\star, \rm init}$ at $z=0$: 

\begin{equation}\label{eq: strip contribution}
\centering
f_{\rm strip} =  \log10\left(\frac{M_{\star,\rm init,max}}{M_{\star, \rm init}}\right)  
\end{equation}

\begin{figure}
	\includegraphics[width=0.45\textwidth]{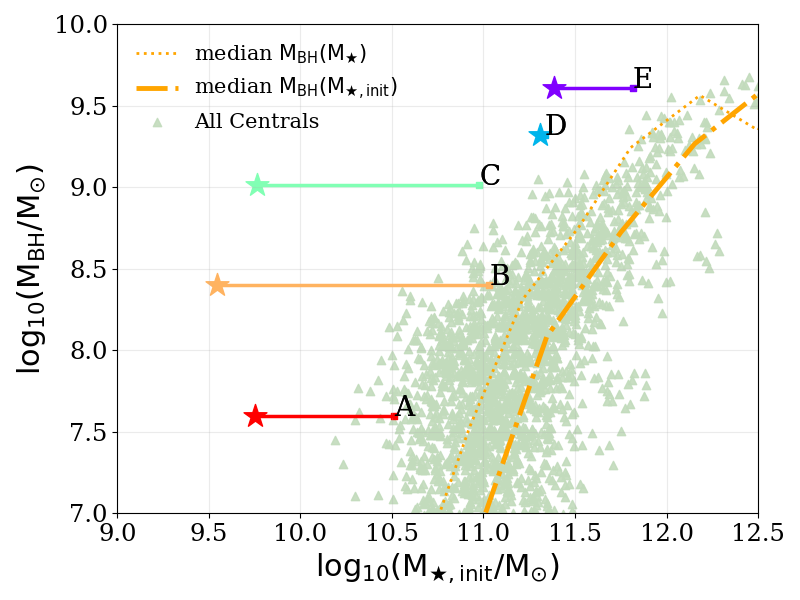}
\caption{Quantification of stellar mass stripping of BMGs. For each BMG, a solid coloured line is drawn between the zero age main sequence stellar mass $M_{\star,\rm init}$ (star symbol) and the maximum $M_{\star,\rm init}$ the galaxy has reached throughout the simulation (square marker), both values correspond to their $M_{\rm BH}$ at $z = 0$. The lengths of the lines correspond to $f_{\rm strip}$ (equation \ref{eq: strip contribution}). The $z=0$ $M_{\rm BH} - M_{\star,\rm init}$ distribution for central galaxies is shown in the background. The dotted orange line represents the median of this distribution, while the dash-dotted orange line represents the median with respect to $\Mstar$ (i.e. including stellar mass loss). From this it appears that stripping is a significant contributing factor in the formation of BMGs. Even so, galaxies C, D and E remain strong outliers when stripping effects are removed.} \label{fig: maxMstar}
\end{figure}

The values of $f_{\rm strip}$ are shown in Table \ref{tab: extreme outliers} for all of our BMGs and are also visualized by the coloured lines in Figure \ref{fig: maxMstar}. The BMGs with the largest $f_{\rm strip}$ (IDs B and C in Table \ref{tab: extreme outliers}) have lost an order of magnitude in $M_{\star}$. \footnote{BMG A would not be considered an outlier when evaluated at its initial stellar mass at $z = 0$. Though it was pushed to the edge of our outlier definition by tidal stripping of its $\Mstar$, it was a additional loss in  $\Mstar$ due to stellar evolution that made it an outlier.} The horizontal movement along the $M_{\star}$ plane in Figure \ref{fig: maxMstar} shows a consistent result to \citetalias{Barber2016}: most of the BMGs are heavily affected by tidal stripping. However galaxies C, D and E are still considered outliers (according to our $z = 0$ definition of being an outlier), even after removing the effect of tidal stripping (i.e. considering their maximum $M_{\star, \rm init}$).

\citetalias{Barber2016} found that a secondary factor causing galaxies to become outliers of the $\MBH(\Mstar)$ relation is their stellar formation time. At earlier times, the normalization of the $M_{\rm BH}(M_{\star})$ relation was higher. This difference in normalization originates from the difference in the density of the universe which causes galaxies at fixed $\Mstar$ to have deeper potential wells \citep{RapidGrowthPhase}. 

When star formation no longer provides enough feedback to regulate the infall of circumgalactic gas onto the galaxy, the central BH enters a rapid growth phase. Black holes have a theoretical maximum accretion rate (the Eddington rate) which scales as $\dot{M}_{\rm BH} \propto \MBH$. However, in practice gas densities are low enough that BH accretion proceeds according to the Bondi-Hoyle rate, which scales as $\dot{M}_{\rm BH} \propto \MBH^2$, which causes a highly non-linear accretion phase. 
The final $\MBH$ depends on the binding energy of the halo, thus at higher $z$ the BH has to grow to a higher mass before it becomes self-regulating \citep{BoothSchaye2010,BoothSchaye2011}.  Isolated galaxies that formed at high $z$ therefore become outliers at $z = 0$, while the rest of the population lowers the normalization. \cite{McAlpine2018} also confirm that the more massive the BH is today, the earlier it began its rapid growth phase (typically $z \approx 6$ for $\MBH$ $10^{9} \Msun$). Note that our definition of an outlier exclusively refers to $z = 0$: our $1\%$ cut-off is not redshift-dependent.  

\begin{figure}
	\includegraphics[trim=0.5cm 0.25cm 0.25cm 0.25cm, clip=true, width=0.45\textwidth]{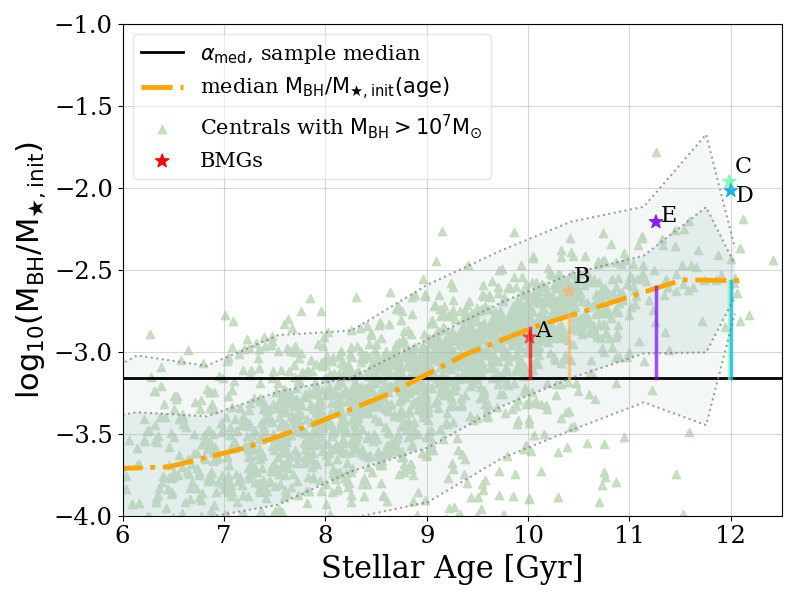} 
\caption{The $\MBH/\rm M_{\star, init}$ ratio as a function of initial stellar mass-weighted ages for central galaxies with $\MBH > 10^{7} \Msun$ at $z = 0$. The dash-dotted orange line shows the median relation, while the solid black line represents the median $\MBH/\rm M_{\star, init}$ value of this entire selection of galaxies. The BMGs are plotted at the maximum $\rm M_{\star, init}$ they ever had, in order to remove the effects of tidal stripping. The tops of the coloured, vertical lines correspond to the value of the median $\MBH/\rm M_{\star, init}$ at the stellar ages of the BMGs. The light and dark gray shaded areas represent the 1 and 2-$\sigma$ standard deviations from the median relation, respectively. It can be seen that, in particular, outliers C, D and E were formed at relatively early times.  \label{fig: age}}
\end{figure}

To test the effect of formation time on our 5 BMGs, we plot in Figure \ref{fig: age} the $M_{\rm BH}/M_{\star,\rm init}$ ratio against the mean stellar ages (weighted by their initial stellar mass) of central galaxies with $M_{\rm BH} > 10^{7} \Msun$. Here we assume that the majority of central galaxies at $z = 0$ have not been affected significantly by stellar stripping and only include galaxies with $M_{\rm BH} > 10^{7} \Msun$, where most galaxies have already been through the rapid BH growth phase, as seen in Figure \ref{fig: dist}. The BMGs are plotted using their maximum $M_{\rm \star,\rm init}$ to exclude stripping effects. The median of the $\MBH/M_{\star,\rm init} - {\rm age}$ relation (black dashed line) is a measure of the contribution that the age of a galaxy has on the scatter in the $\MBH-\Mstar$ relation. Here we see a clear positive trend of $M_{\rm BH}/M_{\star}$ ratio with mean stellar age. This trend represents the evolution of the normalization of the $M_{\rm BH}(M_{\star})$ relation over time.  

As in \citetalias{Barber2016}, we quantify the contribution of age in a similar manner to that of the tidal stripping contribution.
\begin{equation}\label{eq: age contribution} 
\centering
f_{\rm age} = \log10 \left[ \rm med \left( \frac{ \MBH }{ M_{\star, init} } \right)_{age,i} \right]  - \alpha_{\rm med} .
\end{equation}

For a given BMG, $f_{\rm age}$ is the logarithmic difference between the value of the median $M_{\rm BH}/M_{\rm \star, init,max}$ - age relation measured at the stellar age of the galaxy (the black dashed line in Figure \ref{fig: age}) and $\alpha_{\rm med}$ = -3.15, the median $M_{\rm BH}/M_{\star, init}$ value measured for the whole population of central galaxies with $M_{\rm BH} > 10^{7} \Msun$ (the dashed red line in Figure \ref{fig: age}).
The values of $f_{\rm age}$ are shown in Table \ref{tab: extreme outliers} for all five of the BMGs. They are illustrated as vertical coloured lines in Figure \ref{fig: age}, which can easily be compared to the stripping effects for each of these BMGs (horizontal coloured lines) in Figure \ref{fig: maxMstar}. From Figures \ref{fig: maxMstar} and \ref{fig: age}, it can be seen that those BMGs (C, D and E) which could not be explained by tidal stripping have formed at relatively early times. This shows that the effects found by \citetalias{Barber2016} also apply for to the origin of $M_{\rm BH}(M_{\star})$ outliers with $M_{\rm BH} > 10^{9} \Msun$. However, these BMGs still lie on the outer edges of the distribution (at $> 1 \sigma$ from the median relation), even after the effects of both stripping and high stellar age are accounted for. In the next section we study these objects in more detail to better understand their origin. 

\subsection{Detailed analysis of selected BMGs}
\label{ss: detailed analysis}
The formation and evolution of the 5 BMGs (selected based on their resemblance to observational outliers, see Section \ref{sss: BMG selection} for the full definition) were investigated in more detail by studying their morphology and environments through visual inspection of the density of stellar, and BH particles throughout the simulation. An example of such a stellar density profile is shown in Appendix \ref{AP: galaxy properties} where we depict the  outlier ID E at redshifts $z = 0.68, 0.57$ and $0.00$. 

The properties of the BMGs are summarized in Table \ref{tab: extreme outliers}. The majority of these galaxies have lost a significant fraction of their maximum stellar mass due to tidal stripping. 
BMG IDs A and B lie $< 100 $ kpc from their nearest more massive neighbour at $z=0$. They have crossed above the $M_{\rm BH}/M_{\star} > 0.01$ threshold between $z=0.036$ and $z=0$, which leads us to believe that they are still in the process of being tidally stripped. Their stellar density plots and $f_{\rm strip}$ values (0.76 and 1.49 respectively) bot confirm that BMGs A and B are remnants of tidal stripping. 

The BMGs with highest $M_{\rm BH}$ are of particular interest as they have similar $\MBH$ to observed outliers of interest, since the extraordinarily high $M_{\rm BH}/M_{\star}$ ratios of these observed galaxies are still debated \citep{ExplainReportedlyOvermassive,ExplainNGC1271}. The early formation time dominantly contributes to the outlier origins of BMGs D and E. For BMG C, it is tidal stripping that is the dominant process, although its early formation time also strongly contributes to its outlier origin. 
However, after removing the effects of stripping and age, the $M_{\rm BH}/\Mstar$ ratios of BMGs are still on the outskirts of the distribution (see Figure \ref{fig: age}). We will now discuss the origin of these three BMGs in detail.

BMG-D, in contrast to BMGs A and B, has barely been influenced by tidal stripping, which is reflected in its value of $f_{\rm strip} = 0.03$ (see also Figures \ref{fig: evolution} and \ref{fig: maxMstar}). 
This galaxy hosts a very old BH that entered the rapid growth phase directly following a major merger event at $z = 5$, reaching $\MBH$ $> 10^7 \Msun$ by $z = 4$. 
It enters a second rapid growth phase after another major merger event at $z\approx 2.5 $, causing $\MBH$ to increase to $\approx 2 \times 10^9$ $M_{\sun}$, within less than 125 Myr. This triggers an AGN feedback blast wave that removes all the gas from the galaxy's environment, thereby quenching star formation (around $z = 2$). The galaxy resides on the outskirts of a cluster. It has passed through this cluster around $z = 0.6$, at which point it crossed the $\MBH/\Mstar > 0.01$ threshold. At $z = 0$, the galaxy has relaxed and it is left as an isolated massive galaxy. We therefore classify it as a relic of the high-redshift ($z > 2$) Hydrangea Universe. 

Similar to BMG D, BMG C hosts a very old central BH. This central BH entered the rapid growth phase between $z =6.7$ and  $z =5.5$, immediately following a merger event involving two galaxies with near equal $\Mstar$. By $z = 3$ it has reached $\MBH \approx 10^{8.9}$ M$_{\sun}$. It continues to grow in $\MBH$ via gas accretion and a second major merger event at $z = 2$, reaching its maximum $\MBH$ of $ 10^{9.0} \Msun$ by $z\approx 1.9$. 
At redshift $z = 0.6$ it encounters a Brightest Cluster Galaxy (BCG), which starts to strip its stellar mass. This stripping continues up to $z=0$, where we see a clear ‘nugget' stellar core on the outskirts of the BCG. We therefore classify this galaxy as a severely stripped relic of the high-redshift ($z > 2$) Universe. 

The most notable of the BMGs is outlier ID E, since it is not only the BMG with the highest $M_{\rm BH}$ but it is also the only BMG that is classified as a central galaxy at $z=0$. 
Like BMGs C and D, it contains an old central BH. An intermediate merger at $ z = 2.7$ triggers the rapid growth phase of its central BH, increasing its black hole mass to $\MBH = 10^{9.45} \Msun$. The corresponding AGN feedback blast has cleared all the gas from the galaxy by $z = 2.0$, and the galaxy passes the $\MBH/\Mstar > 0.01$ limit around the same redshift. A third and fourth intermediate merger occur at $ z = 2.3$ and $z = 1.3$, where the merging companions host black holes of $\MBH \approx 10^{7.4} \Msun$ and $\MBH = 10^{7.9} \Msun$ respectively. The latter two mergers increase the initial stellar mass to  $\Mstar = 10^{11.5} \Msun$ and $\Mstar = 10^{11.7} \Msun$ respectively. These merging events are visible in Figure \ref{fig: evolution} as the `staircase' shape that BMG E (solid red line) follows at high $\MBH$.

From $z = 0.7$ onwards, BMG E interacts with a nearby proto-cluster, which causes the outer stellar halo to completely disconnect from the galaxy-core (the central stars and BH) at $z \approx 0.6$. This event decreased its half-mass radius by $83\%$ and causes BMG E to lose 65\% of its stellar mass (this event is shown in the left-hand and middle panels of Figure \ref{fig: Outlier E}). The outer halo, instead of being shredded, continues to exist as a separate, core-less subhalo. 
The central BH and the stellar core remain as a compact, fairly isolated object that escapes from this proto-cluster and has time to re-adjust towards an equilibrium state such that it does not appear particularly disturbed at $z=0$ (see the right-hand panel of Figure \ref{fig: Outlier E}). Although this would be considered an unprobable event, it is not necessarily unphysical.  
We thus conclude that although this galaxy is an isolated (with its nearest more massive neighbouring subhalo at a distance of > 1.15 Mpc) central at $z = 0$, it has experienced significant environmental influence as a satellite in the past. Similar to BMG C, this galaxy is therefore classified as a severely stripped relic of the high-redshift ($z > 2$) Universe. 


BMG C, D and E became outliers immediately after their BH went through the rapid growth phase at very high redshift ($z \gtrsim 2$; see Table \ref{tab: extreme outliers}). All three of these galaxies have furthermore experienced multiple merger events at high redshift ($z > 2$). Since these BMGs entered the rapid growth phase within 500 Myrs of a merger event, it is possible that these merger events created initial conditions of the rapid growth phase that drove their $\MBH$ to anomalously high values.
\cite{McAlpine2018} predict that galaxy interactions, while relevant at low $0<z<2$, are less important for triggering the rapid growth phase at high redshifts ($z > 2$). \cite{Steinborn2018} investigated the effect of major mergers on the AGN luminosity in the Magneticum Pathfinder simulations. They found that $>50\%$ of the luminous AGN ($L_{\rm AGN} > 10^{45}$ erg/s) at $z \approx 2$ have experienced a merger within the last 0.5 Gyr. However, they argue that, to some extent, this result reflects the intrinsically high merger rates of massive galaxies in which luminous galaxies typically reside. \cite{Steinborn2018} furthermore found that although mergers increase the probability for nuclear activity by a factor of three, they still play only a minor role ($<20 \%$) in causing nuclear activity in the overall AGN population.
These major merger events immediately preceding the rapid growth phase at $z > 2$ would thus be considered rare, or at least uncommon coincidences. It is therefore plausible that if major mergers fuelled the central BHs of BMGs C, D and E, this drove them to anomalously high BH masses, which in turn causes them to still be offset in Figure \ref{fig: age}.

\subsection{Comparison to observations}
\label{sss: Comparison to observations}
\begin{figure}
	\includegraphics[trim=0.5cm 0.25cm 0.25cm 0.25cm, clip=true, width=0.45\textwidth]{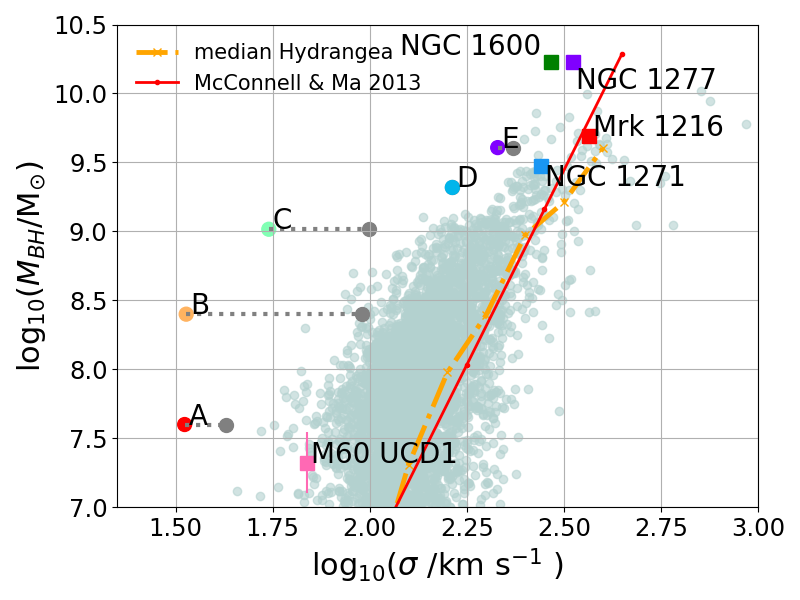} 
\caption{$\MBH$ as a function of stellar velocity dispersion. The distribution for galaxies with log$_{10}$($\MBH$/$\Msun$) > 7 at $z = 0$ is shown in gray-blue. The solid red line shows the empirical relation for early type galaxies from \protect\cite{McConnellMa2013}, while the dash-dotted orange line displays the median of the Hydrangea sample. The Hydrangea BMGs are plotted at their $z = 0$ values (coloured circles). The re-computed stellar velocity dispersions are shown as gray circles connected to the original values by dotted gray horizontal lines. \label{fig: MbhSigma}}
\end{figure}

\begin{figure}
	\includegraphics[trim=0.5cm 0.25cm 0.25cm 0.25cm, clip=true, width=0.45\textwidth]{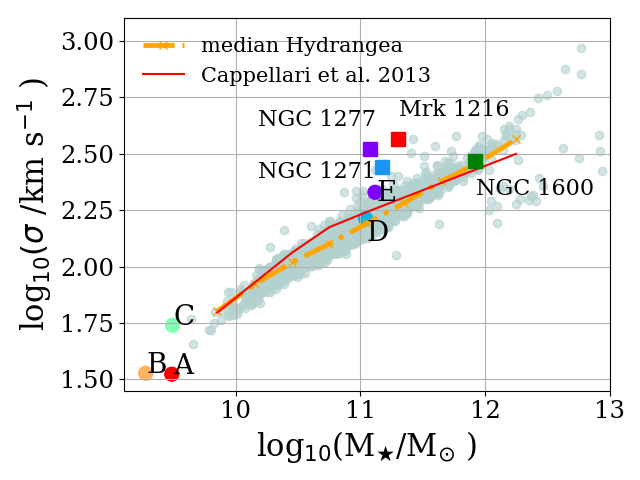} 
\caption{Stellar velocity dispersion as a function of $\Mstar$. The distribution for galaxies with log$_{10}$($\MBH$/$\Msun$) > 7, $\sigma$> 100 kms$^{-1}$ and sSFR > 0.1 $\Msun$/Gyr  at $z = 0$ is shown in gray-blue. The empirical relation from \protect\cite{Cappellari2013} is shown as a solid red line, and the median relation of the Hydrangea sample as a dash-dotted orange line. Observed and simulated BMGs are plotted in the same styles as Fig 5. BMGs D and E are not strong outliers above the sigma-Mstar relation.\label{fig: mstar_sigma}}
\end{figure}

It is interesting to know whether the origins of the simulated BMGs could be used to explain the origins of observed $\MBH(\Mstar)$ outliers. Compact positive outliers from the $\MBH-\Mstar$ relation are observationally characterized by following the $\MBH(\sigma)$ relation, and by being offset from the Faber-Jackson relation \citep{ObsNGC1271,ObsNGC1277}.
In order to facilitate a fair comparison to observational outliers, we first evaluate how well the simulations reproduce the empirical $\MBH(\sigma)$ and Faber-Jackson relations (see Figures \ref{fig: MbhSigma} and \ref{fig: mstar_sigma}). Since these relations could depend sensitively on how either property is measured, observed outliers should only be compared to their respective empirical relation, and likewise, simulated BMGs to their respective relation from the Hydrangea sample.\\

In the Hydrangea simulation, the stellar velocity dispersion is defined as $\sigma = \sqrt[]{2 E_{kin,\star}/3 M_{\star}}$ where $E_{kin, \star}$ is the kinetic energy of stars, and $M_{\star}$ is the total mass of the stellar particles. Both $E_{kin, \star}$ and $M_{\star}$ were summed over all the stellar particles that are part of the subhalo, as identified by the SUBFIND algorithm. This one dimensional velocity dispersion implicitly assumes an isotropic stellar velocity dispersion.
From Figure \ref{fig: MbhSigma}, it is immediately apparent that the velocity dispersion of BMGs A, B and C are remarkably low. Evaluating their $\sigma$ values over the course of the simulation showed that these galaxies had an order of magnitude higher velocity dispersion in the past. 
For BMG A and B, this drop in velocity dispersion occurs simultaneously to the tidal stripping events described in subsection \ref{ss: detailed analysis}. This may indicate that the high-velocity stars have been stripped from A and B. As discussed in section \ref{ss: detailed analysis}, these two galaxies are most likely still in the process of being tidally stripped, which implies that they are not in equilibrium at $z=0$.
Furthermore, if the velocity dispersions of the BMGs were to be measured in a more observationally consistent way, their values could increase. This is especially relevant for BMGs A, B and C, since these galaxies reside in a dense environment ($< 60$ pkpc from their host), and an observational aperture measurement will include more high-velocity background particles.
To test the latter hypothesis, the velocity dispersions were re-computed within a spherical aperture of 5 kpc (including all star particles irrespective of their subhalo membership). With this, the velocity dispersions are indeed higher at $z = 0$ (43 kms$^{-1}$, 95 kms$^{-1}$ and 100 kms$^{-1}$ for A, B and C respectively). As can be seen in Figure \ref{fig: MbhSigma}, when using the re-computed stellar velocity dispersions, BMGs A, B and C lie closer to the median of the Hydrangea sample. For BMGs D and E, recomputing the velocity dispersions within a 5 kpc aperture made no significant difference. A velocity dispersion measurement using a cylindrical aperture was also tested. This increased the velocity dispersions by another $\sim 50\%$.
This implies that the velocity dispersions of NGC 1271 and NGC 1277, which both reside in a dense cluster environment, could also be biased high due to the observations being unable to distinguish between host and satellite.

In Figure \ref{fig: MbhSigma}, it can be seen that the median relation from the Hydrangea sample agrees with the empirical $\MBH(\sigma)$ relation from \cite{McConnellMa2013} (the dash-dotted orange and solid red lines respectively). 
While NGC 1271 and Mrk 1216 lie along the empirical relation from \cite{McConnellMa2013}, NGC 1277, NGC1600 and M60 UCD1 are outliers above the $\MBH(\sigma)$ relation.
For M60 UCD1, the distance above the $\MBH(\sigma)$ relation is comparable to the number of dex that BMGs B and C lie above the median relation of Hydrangea. 
However, we caution that the velocity dispersions of A, B and C lie outside of the observed range of velocity dispersions from \cite{McConnellMa2013}, thereby ruling out definitive comparisons.
NGC 1277 and NGC 1600 respectively lie 0.66 and 0.98 dex above the $\MBH(\sigma)$ relation from \cite{McConnellMa2013}, which is comparable to the offsets of BMGs D and E from the median relation in Hydrangea (1.15 and 1.02 dex, respectively).   

The luminosities needed for the computation of the Faber-Jackson relation of the Hydrangea sample require stellar population synthesis modelling of the stellar particles, which is beyond the scope of this work. However, to first order, luminosity is proportional to $\Mstar$. In Figure \ref{fig: mstar_sigma}, we therefore compare the BMGs to the observed $\sigma(\Mstar)$ relation for early type galaxies \citep[from][]{Cappellari2013}. 
The $\sigma(\Mstar)$ relation measured by \cite{Cappellari2013} agrees well with the median value of $\sigma(\Mstar)$ for the whole observed $\Mstar$ range (the solid red and dash-dotted orange lines in Figure \ref{fig: mstar_sigma}).
Simulated BMGs A, B and C lie outside the domain of this relation.
The observed outliers NGC 1277 and Mrk 1216 are positive outliers ($0.27$ dex each) to the relation from \cite{Cappellari2013}. Assuming that these observed $\sigma$ values are correct, this indicates that they either have a DM halo which is much more massive than typical for their $\Mstar$, or that they are out of equilibrium. I.e., they have very recently been stripped dramatically, but the (remaining) stars have not yet responded to the shallower potential well. The former explanation seems very unlikely for NGC 1277, since we expect satellites to have less DM at fixed $\Mstar$. The latter explanation, is in line with what we found for BMG E (see section \ref{ss: tidal stripping formation time}). 
Simulated BMGs D and E have a small offset from the median of the Hydrangea sample (0.02 and 0.12 dex above the median respectively), which is comparable to the offset of NGC 1600 and NGC 1271 to the empirical relation (0.04 and 0.17 dex respectively).
BMGs D and E are very compact galaxies (with $R_{e} =3.93$ kpc and $R_{e} = 2.61$ kpc). It could be that the stellar velocity dispersions of very compact galaxies are underestimated due to numerical resolution effects, since at lower resolution, the softening artificially smooths out the density in the centre, while at higher resolution, the central potential well should become deeper, and so the velocity dispersion of the galaxy, in the very centre, might increase.

Figures \ref{fig: MbhSigma} and \ref{fig: mstar_sigma} show that the $\MBH(\sigma)$ and $\sigma(\Mstar)$ relations from Hydrangea are comparable to empirical results within the appropriate $\sigma$ and $\Mstar$ ranges. Hence, BMGs D and E could in particular be relevant for the understanding of the origins of observed compact galaxies with high $\MBH/\Mstar$.
In terms of $\MBH$ and $\Mstar$, BMGs D and E are comparable to the observed outliers NGC 1271, \citep{ObsNGC1271}, NGC 1277 \citep{NGC1277.Bosch2012,ObsNGC1277} and Mrk 1216 \citep{YildirimMrk1216Yr2015, 2017BHmassMrk1216}.
There are no outliers with $\MBH > 10^{10} \Msun$, which is similar to the inferred $\MBH$ of NGC 1600 \citep[$\MBH = 1.7  \pm 0.15 \times 10^{10} \rm M_{\sun}$,][]{ObsNGC1600}. This can be explained by the general lack of galaxies in this mass range (see Figure \ref{fig: dist}). A simulation with even better statistics at the high-mass end would be required to investigate such systems.\\

We will now compare the BMGs D and E to observed outliers with similar $\MBH$ and $\Mstar$.

The BMGs D and E have one-dimensional stellar velocity dispersions $\sigma$ = 163 and  213 km s$^{-1}$, respectively, while the observed outliers NGC 1271, NGC1277 and Mrk 1216 have $\sigma$ = $276^{+73}_{-4}$, $333 \pm 12$ and $368 \pm 3$ km s$^{-1}$, respectively (see Table \ref{tab: extreme outliers}). The quoted stellar velocity dispersions of NGC 1271 and 1277 specifically exclude the sphere of influence of the black hole. However, this is not the case for Mrk 1216, where $\sigma$ is defined as the average stellar velocity dispersion within the effective radius. 

As discussed above, it could be that the stellar velocity dispersion is underestimated for galaxies as compact as our BMGs, due to numerical resolution effects.
However, the simulated $\sigma$ values of BMGs D and E lie within the spread of the velocity dispersion at fixed $\Mstar \approx 10^{11} \Msun$ from the ATLAS$^{\rm 3D}$ sample \citep{Cappellari2013}. According to \cite{Cappellari2013}, M $\propto \sigma^{2} \times R_{e}$, where $R_{e}$ is the projected half-light radius. This implies that at a constant $\Mstar$, the velocity dispersion depends mainly on the size of the galaxy. The observed values of $R_{e}$ for NGC 1271 ($R_{e}$ = $2.2 \pm 0.9$ kpc) and NGC 1277 ($R_{e}$ = $1.2 \pm 0.1$ kpc), are two to three times lower than the half-mass radii of BMGs D ($R_{e}$ = 3.93 kpc) and E ($R_{e}$ = 2.61 kpc). This difference in size may account for the lower velocity dispersions in our simulated BMGs relative to these observed examples, despite having very similar $\Mstar$. 
\newline The ages of both NGC 1277 and NGC 1271 are estimated to be greater than 10 Gyr \citep{FerreMateu2015}, which is consistent with the ages of all of our simulated BCGs. The age of Mrk 1216 is even estimated to be >12 Gyr \citep{Mrk1216andPGC032873from2017}. This is comparable to BMG D, which is primarily an outlier due to its early formation time.
\newline Another observed characteristic of compact outliers is their fast-rotator structure \citep{YildirimMrk1216Yr2015, ObsNGC1271,ObsNGC1277, Mrk1216andPGC032873from2017}.
From the analysis of \cite{Lagos2018}, BMG E is a slow rotator ($\lambda_{R}$ close to 0), which adds to the potential differences in detail discussed above.
\newline All of the BMGs in the simulation reside in a very rich cluster environment. It is thus not surprising that most outliers have been heavily stripped during their lifetimes. 
NGC 1271 and NGC 1277 are both embedded in the nearby Perseus galaxy cluster and are relatively compact galaxies \citep[effective radius smaller than 2.5 kpc][]{ObsNGC1277,ObsNGC1271, NGC1277Clust2014}, however they do not show any obvious signs of having been stripped. 
Mrk 1216 resides in  a significantly different environment; it has only two neighbouring galaxies within 1 Mpc \citep{YildirimMrk1216Yr2015}. This galaxy is also relatively compact, though deep field imaging again show no apparent signatures of interaction or stripping \citep{Mrk1216andPGC032873from2017}.

As discussed above, BMGs D and E share characteristics with the observed outliers NGC 1271, NGC 1277 and Mrk 1216, but also differ in some details such as their offsets from the $\MBH$($\sigma$) and $\sigma$($\Mstar$) relations or kinematic structure (fast vs. slow rotator). It is unclear whether these differences are related to their BMG nature, or reflect unrelated shortcomings of the simulation model. Further, detailed work would be required to investigate this issue further (see also \citealp {vandeSande2018} and \citealp{Lagos2018}). \\

The similarities described above, between NGC 1271, NGC 1277, Mrk 1216 and the simulated BMGs D and E could hint towards a similar origin: relics of the $z > 2$ Universe. 
It is important to note that these `high-z relics' may also have been significantly stripped, as is the case for both BMG C and E, which have high values for both $f_{\rm strip}$ and $f_{\rm age}$. The fact that the nearest more massive neighbour of BMG E lies at > 1 Mpc (which is the case for Mrk 1216) does not necessarily imply that they have not undergone tidal stripping at earlier times, as they may have drifted away from these more massive neighbours. NGC 1271, NGC 1277 and Mrk 1216 may also have gone through a similar tidal stripping period but `survived' to eventually end up seemingly undisturbed at $z = 0$. It is therefore interesting to study the environments of  outliers out to scales of > 1 Mpc, and we encourage observational efforts to do so.\\

\subsection{Alpha enhancement}
\label{ss: Mg/Fe}
The [Mg/Fe] ratio is an indicator of how extended the SF period of a galaxy has been, since core-collapse supernovae (SNe), which take place shortly after a star formation episode, release $\alpha$-elements (such as the easily observable Mg), while Type Ia SN, which occur mostly after $\sim 1$ Gyr, release primarily iron (Fe) and no $\alpha$-elements. The more abruptly SF is quenched, the less time there is to form Fe via Type Ia SNe and to incorporate it into stars. \cite{MarijkeAlphaFe} have shown that in EAGLE, AGN feedback is responsible for its success in reproducing the observed $\alpha$-enhancement of massive galaxies. We thus expect a positive correlation between [Mg/Fe] and $M_{\rm BH}$ at fixed $\Mstar$. Recently, \cite{HGFE} observed that positive galaxy outliers of the $M_{\rm BH}-\sigma$ relation are older and more $\alpha$-enriched. Because of the similarity between the observed $M_{\rm BH}-\sigma$ and the $M_{\rm BH}-M_{\star}$ relations, we have examined these relations in the Hydrangea simulations. 

To examine the secondary trends in [Mg/Fe] and $\MBH$, it is important that the simulated [Mg/Fe]($\sigma$) relation matches observations. \cite{Segers2016} and \cite{MarijkeAlphaFe} showed that the EAGLE simulated trends between [$\alpha$/Fe], galaxy stellar mass and age are in good agreement with observations of early-type galaxies for galaxies with log$_{10}$($\Mstar$/$\Msun$)> 10.5.
Furthermore, when we examine the distribution of simulated galaxies with $\sigma$ > 100 km s$^{-1}$ and sSFR > 0.1 $\Msun$/Gyr (in order to facilitate comparison with the observational data) it appears that BMGs D and E have relatively high [Mg/Fe] values at fixed $\sigma$, which is in qualitative agreement to the observations.
This, in combination with the overlap between the $\MBH(\sigma)$ relation in Hydrangea, and the empirical relation of \cite{McConnellMa2013} for log$_{10}$($\sigma$/kms$^{-1}$) > 2.2 (see Figure \ref{fig: MbhSigma}), argue for a meaningful comparison of secondary trends in [Mg/Fe] and $\MBH$ between the Hydrangea simulation and observations. 

\begin{figure}
   	\includegraphics[width=0.49\textwidth]{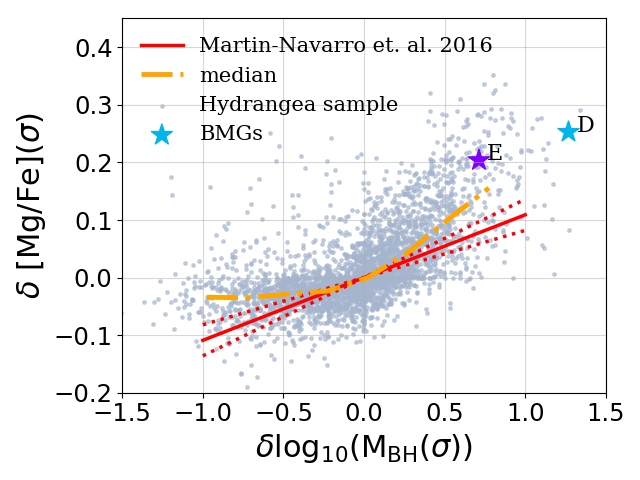} 
\caption{Residuals of the median [Mg/Fe]$(\sigma)$ relation as a function of residuals of the median $M_{\rm BH}(\sigma)$ relation. The sample from Hydrangea includes only galaxies with $\sigma > 100$ km s$^{-1}$, sSFR > 0.1 $\Msun$/Gyr and $\MBH > 10^7 \Msun$, and is plotted in purple-gray. The BMGs fulfilling the former requirement are shown as star symbols. The median of the Hydrangea sample is plotted as a dash-dotted orange line. The observed relation from \protect\cite{HGFE} is shown as a red solid line, with the red dotted lines showing the uncertainty in the fit. Galaxies with higher $M_{\rm BH}$ tend to have a higher [Mg/Fe] ratio at fixed $\sigma$, in broad agreement with the observational results of \protect\cite{HGFE}. \label{fig: Mg/Fe}} 
\end{figure}

In Figure \ref{fig: Mg/Fe} we plot the residuals relative to the median [Mg/Fe]$(\sigma)$ relation as a function of the residuals of the median $M_{\rm BH}(\sigma)$ relation for galaxies in the Hydrangea simulations at $z=0$ with $\sigma > 100$ km s$^{-1}$, sSFR > 0.1 $\Msun$/Gyr and $\MBH > 10^7 \Msun$, where the first two constraints facilitate comparison to the observational data and the latter avoids resolution effects. 
In this way, we can investigate the relation between $M_{\rm BH}$ and the SF history at fixed $\sigma$. 
The observed relation from \cite{HGFE} is shown in red, with the dashed lines showing the uncertainty in the fit. 

It can be seen that the scatter in $M_{\rm BH}(\sigma)$ correlates positively with the scatter in [Mg/Fe]($\sigma$), in qualitative agreement with the findings of \cite{HGFE}. The difference in functional form may be due to selection effects, which are also reflected in the sample sizes; \cite{HGFE} uses a very heterogeneous sample of 57 galaxies, while we include $\sim 3400$ galaxies. The vast majority of the galaxy sample from \cite{HGFE} resides in the domain -0.5 < $\delta \rm log_{10}(\MBH(\sigma))$ < 0.5, which is also the domain in which the median relations are in best agreement.

Although only 2 out of 5 BMGs have $\sigma > 100$ km s$^{-1}$ at $z=0$, they appear to be in line with the positive correlation between the scatter in $M_{\rm BH}(\sigma)$ and the scatter in [Mg/Fe]($\sigma$). The [Mg/Fe] abundance of outlier D ([Mg/Fe] = 0.36) and E ([Mg/Fe] = 0.31) are similar to the best-fitting value from \cite{HGFE} for NGC 1277 ([Mg/Fe] = 0.32) and Mrk 1216 (0.26 $\pm$ 0.05 \citealt{Mrk1216andPGC032873from2017}).  Note that uncertainties in stellar population modelling lead to systematic errors in observed alpha-enhancement values of $\approx 0.1$ dex, while nucleosynthetic yields (used in the simulations) are uncertain by a factor $\approx 2$ \citep{Wiersma2009b}. Thus, BMGs D and E have [Mg/Fe] consistent with the observed outliers. This points towards a similar SF history.

\section{Summary}
\label{sec: conclusion}
We have used the Hydrangea simulations \citep{Hydrangea}, part of the C-EAGLE project \citep{C-EAGLE}, of massive galaxy clusters and their surroundings to study the evolution and origin of galaxies with central SMBHs that are overmassive relative to the general $M_{\rm BH} - M_{\star}$ relation at $z = 0$ in galaxy cluster environments. This work builds on \citetalias{Barber2016}, who used the (100 Mpc)$^{3}$ EAGLE simulation to study the origin of overmassive black holes with smaller masses and in less overdense environments than studied here. 
We probe galaxies in a mass regime comparable to observed $M_{\rm BH}(M_{\star})$ outliers. We define $M_{\rm BH}(M_{\star})$ outliers within the simulation as those with $M_{\rm BH}/M_{\star} > 0.01 $ and $M_{\rm BH} > 10^{7} \Msun$. \\

From this initial set of $M_{\rm BH}(M_{\star})$ outliers, a selection by eye of the five most extreme outliers of this relation, which we term ``Black hole monster galaxies'' (BMGs), has been made based on their location in the $M_{\rm BH} - M_{\star}$ plane (Figure \ref{fig: dist}). They were chosen such that they have masses similar to observed galaxies that are extreme positive outliers of the observed $\MBH-\Mstar$ relation. We have traced the BMGs through time to study their formation and evolution. Our main findings are as follows: 
\begin{itemize}

\item We identified 69 galaxies that fulfil our $\MBH(\Mstar)$ outlier criteria at $z = 0$, of which 77$\%$ are satellites. The outliers lie closer to their host and are significantly more compact than other satellites of similar $\Mstar$, suggesting that tidal stripping is an important contributor toward galaxies becoming outliers. Outliers typically have $M_{\star} = [10^{9} - 10^{11}] \Msun$, where our most massive BMG (ID E, see Table \ref{tab: extreme outliers}), reaches a stellar mass of $1.31 \times 10^{11} \Msun$ and $M_{\rm BH} = 4.05 \times 10^{9} \Msun$ (see Figure \ref{fig: dist}). The simulations therefore reproduce the observed existence of $M_{\rm BH}(M_{\star})$ outliers with $M_{\rm BH} > 10^{9} \Msun$.

\item  In agreement with \citetalias{Barber2016}, our results suggest that the origin of galaxies with such over-massive BHs is primarily due to a combination of tidal stripping \citep[Figure \ref{fig: maxMstar}, see also][]{Volonteri2008, Volonteri2016} and their formation at high redshift (Figure \ref{fig: age}). For the BMGs with $\MBH > 10^{9} \Msun$, galaxy-galaxy merger events also plausibly play a role in the formation of their anomalously high-mass central BHs.

\item  Galaxies with higher $M_{\rm BH}$ tend to have a higher [Mg/Fe] ratio at fixed stellar velocity dispersion, in agreement with the observational results of \cite{HGFE} (Figure \ref{fig: Mg/Fe}). This correlation is thought to underline the strong influence of AGN feedback on the star formation history of high-mass galaxies \citep{MarijkeAlphaFe}.

\item BMGs D and E have similar $\MBH$ and $\Mstar$ as observed $\MBH(\Mstar)$ outlier galaxies NGC 1271 \citep{ObsNGC1271}, NGC 1277 \citep{ObsNGC1277} and Mrk 1216 \cite{2017BHmassMrk1216}. Both D and E host a BH that entered the rapid growth phase at $z>2$ and both exhibit high [Mg/Fe] values, consistent with being ``relics'' of the high-$z$ ($z > 2$) Hydrangea universe.
The properties of these BMGs are summarized in Table \ref{tab: extreme outliers} (see the discussion in Section \ref{ss: detailed analysis}). 
BMG E is a `survivor' of tidal stripping events (see Figure \ref{fig: Outlier E}). Its outer stellar halo was completely disrupted by tidal interactions with a proto-cluster at $z\approx 0.5$, and the remaining core does not appear particularly disturbed at $z=0$. We therefore encourage observational efforts to search for outliers in clusters even out to large ($\sim$1 Mpc) distances, which corresponds to the distance of outlier E to its nearest more-massive neighbour at $z=0$. 
Due to the similarities between the observed outliers and the simulated BMGs, a similar origin is plausible. This means that in addition to being a high-$z$ relic, outliers could very well have been stripped at an earlier time, survived, and appear undisturbed at $z=0$.
On top of this, all BMGs were found to have experienced a merger event immediately preceding the rapid growth phase of their central BH. It is therefore plausible that major merger events triggering the rapid growth phase of BHs could contribute to the formation of anomalously massive SMBHs

\end{itemize} 
 
Based on hydrodynamical, cosmological simulations we thus expect the existence of observed $M_{\rm BH}(M_{\star})$ outliers with $M_{\rm BH} > 10^{9} \Msun$. Observed outliers with such large $\MBH$ are plausibly survivors of tidal stripping and it would be interesting to study their environment out to $\sim$1 Mpc. The positive correlation between the scatter in $M_{\rm BH}(\sigma)$ and in [Mg/Fe]($\sigma$) indicates a strong relation between the $\MBH$ and the time-scale of the quenching process for massive galaxies. It would furthermore be interesting for future studies to investigate the relation between merger-driven rapid growth phases at high redshifts ($z > 2$) and anomalously high-mass SMBHs. 

\section*{Acknowledgements}
This work used the DiRAC Data Centric system at Durham University, operated by the Institute for Computational Cosmology on behalf of the STFC DiRAC HPC Facility. DiRAC is part of the National E-Infrastructure. This equipment was funded by BIS National E-infrastructure capital grant ST/K00042X/1, STFC capital grants ST/H008519/1 and ST/K00087X/1, STFC DiRAC Operations grant ST/K003267/1 and Durham University. STK and DJB acknowledge support from STFC through grant ST/L000768/1. RAC is a Royal Society University Research Fellow. The Hydrangea simulations were in part performed on the German federal maximum performance computer ‘Hazel Hen’ at the maximum performance computing centre Stuttgart (HLRS), under project GCS-HYDA/ID44067 financed through the large-scale project ‘Hydrangea’ of the Gauss Centre for Supercomputing. Further simulations were performed at the Max Planck Computing and Data Facility in Garching, Germany. This project received funding from the EU Horizon 2020 research and innovation programme under Marie Skłodowska-Curie grant agreement 747645 (ClusterGal). This work has furthermore benefited from the use of Py-SPHViewer (Benitez-Llambay 2015). 
\bibliographystyle{mnras}
\bibliography{umgals}

\appendix
\section{BMG properties}
\label{AP: galaxy properties}
In Table \ref{tab: extreme outliers} we list all of the relevant properties of the 5 BMGs and observed outliers. Additionally, we list information on some of the most notable observed $M_{\rm BH}(M_{\star})$ outliers. For NGC 1271 and NGC1277, the [Mg/Fe] values are the best-fit values derived by \cite{HGFE}. In column 7, the `outlier $z$' represents the lower limits of a 500 Myr time step.

\begin{figure*}
    \centering
	\includegraphics[trim=2.2cm 1.0cm 1.0cm 0.5cm, clip=true, width= 0.3\textwidth]{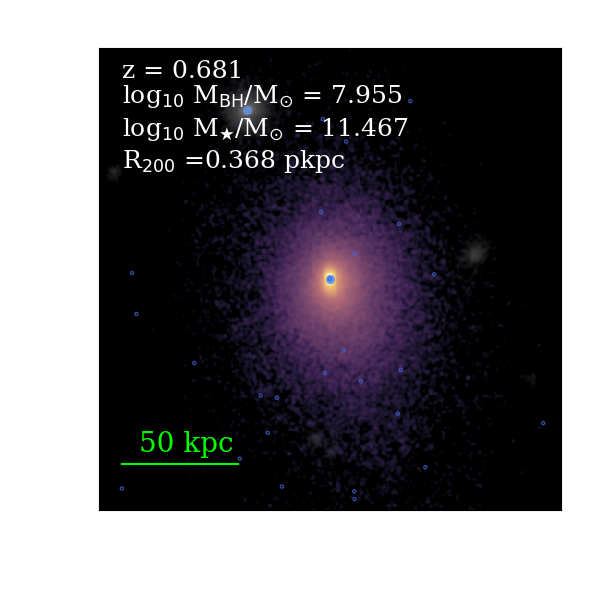}
	\includegraphics[trim=2.2cm 1.0cm 1.0cm 0.5cm, clip=true, width= 0.3\textwidth]{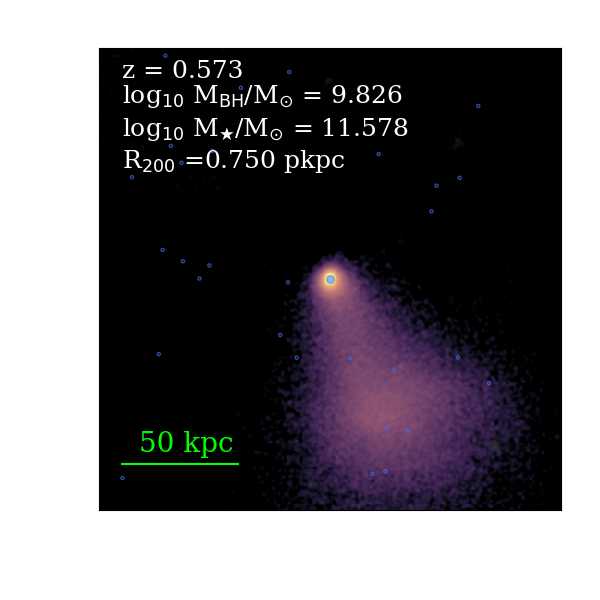}
    \includegraphics[trim=2.2cm 1.0cm 1.0cm 0.5cm, clip=true, width= 0.3\textwidth]{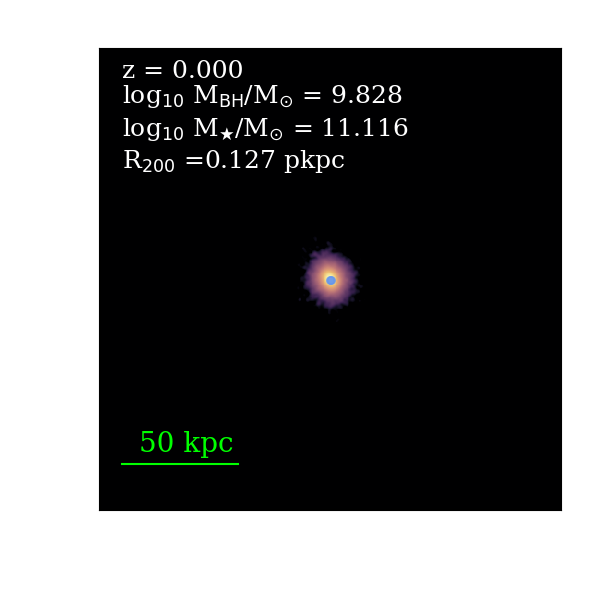}
\caption{Stellar surface density distribution of BMG-ID E at three different snapshots. These images are created with the publicly available Py-SPHViewer, which uses the Smoothed Particle Hydrodynamics (SPH) interpolation scheme. The colours represent the projected density of stellar particles bound to BMG-ID E, with blue/pink corresponding to low density and yellow corresponding to high-mass density. Black hole particles are indicated with blue dots. The most relevant parameters of the galaxy at each snapshot are written on the top left of each image. A green scale bar shows the scale of the images in comoving kpc. The left and middle panels correspond to $z = 0.68$  and $0.57$ respectively, just before and during the stellar stripping event. The right panel depicts the galaxy at $z=0$. This BMG is located in a dense environment right after being stripped. However at $z = 0$ it has `escaped' to a low density region, causing it to be the central galaxy of its FOF group. This portrays an example of an BMG surviving the tidal stripping process. \label{fig: Outlier E}}
\end{figure*}

In Figure \ref{fig: Outlier E} we have plotted the stellar mass density of outlier ID E and its surroundings at three different snapshots throughout the simulation.
The remaining images are available at \url{https://sites.google.com/view/vson-umbh-galaxies/homepage}

\begin{table*}
\caption{The five BMGs and their properties. Column 1 shows the BMG ID (for simulated galaxies) or name (for observations). From left to right: BH mass (column 2), stellar mass (column 3), half-mass radii for the simulated galaxies, or effective radius for observations (column 4), one dimensional stellar velocity dispersion, which is computed from the average of the three dimensional velocity dispersion (column 5), mass-weighted stellar age (column 6), and [Mg/Fe] abundance (column 7). Column 8 shows the redshift of the snapshot at which each galaxy crossed the a $M_{\rm BH}/M_{\star} > 0.01$ threshold. These redshifts have an uncertainty of 500 Myr. Column 9 is the distance to the nearest more massive neighbour of each BMG at $z=0$. The last two columns (10 and 11) show $f_{\rm strip}$ and $f_{\rm age}$ from equations \ref{eq: strip contribution} and \ref{eq: age contribution}, respectively. The last four rows show the observational values for NGC1271, NGC1277 and Mrk 1216 as determined by \protect\cite{ObsNGC1271}, \protect\cite{ObsNGC1277}, \protect\cite{FerreMateu2015}, \protect\cite{Mrk1216andPGC032873from2017} and \protect\cite{HGFE}.  }
\resizebox{\textwidth}{!}{
\begin{tabular}{|c|l|l|l|l|l|l|l|l|l|l}
\hline \hline
\pbox{3cm}{Name\\ID} &  \pbox{3cm}{ log$_{10}$(M$_{\rm BH}/\Msun$) } &  \pbox{3cm}{log$_{10}$($M_{\star}$/$\Msun$) } &  \pbox{3cm}{$R_{e}$ \\ ${\rm [kpc]}$ } & \pbox{3cm}{$\sigma$ \\ ${\rm [km s^{-1}]}$ } & \pbox{3cm}{Age \\ ${\rm [Gyr]}$} &  \pbox{3cm}{[Mg/Fe] } & \pbox{3cm}{Outlier z } & \pbox{3cm}{$D_{\rm host}$ \\ ${\rm [kpc]}$ } & \pbox{3cm}{$f_{\rm strip}$} & \pbox{3cm}{$f_{\rm age}$}  \\ \hline \hline
A & 7.60 & 9.48  & 1.49 & 33  & 10.02 & 0.02  & 0.00 & 25    & 0.76 & 0.30 \\
B & 8.40 & 9.27  & 2.28 & 34  & 9.94  & -0.01 & 0.04 & 30    & 1.49 & 0.38 \\
C & 9.02 & 9.49  & 1.53 & 55  & 10.40 & 0.40  & 2.35 & 60    & 1.22 & 0.59 \\
D & 9.32 & 11.04 & 3.93 & 163 & 11.99 & 0.36  & 0.62 & 475   & 0.03 & 0.59 \\
E & 9.61 & 11.12 & 2.61 & 214 & 10.96 & 0.31  & 2.35 & 1,150 & 0.43 & 0.55 \\ \hline \hline
NGC 1271 & $9.48 \pm 0.15 $ & $11.18 \pm 0.20 $ & $2.2 \pm 0.9$ & 276$^{+73}_{-4}$ & >10            & 0.23            &  &  &  &  \\
NGC 1277 & $9.69 \pm 0.14$  & $11.08 \pm 0.14 $ & $1.2 \pm 0.1$ & 333 $\pm$12      & >10            & 0.32            &  &  &  &  \\
Mrk 1216 & 9.69 $\pm$ 0.15  & 11.30 $\pm$ 0.17  & $2.3 \pm 0.1$ & 368 $\pm$ 3      & 12.8 $\pm$ 1.5 & 0.26 $\pm$ 0.05 &  &  &  &  \\

\end{tabular} }
\label{tab: extreme outliers}
\end{table*}

\bsp	
\label{lastpage}
\end{document}